\documentclass[twocolumn,superscriptaddress,amssymb, nobibnotes, aps, prb]{revtex4-2}
\usepackage{amsmath}
\usepackage{graphicx}
\usepackage{amsfonts}
\usepackage{amssymb}
\usepackage{bm}
\usepackage{dsfont}
\usepackage{epsfig,float,afterpage}
\usepackage[colorlinks,allcolors=blue]{hyperref}
\usepackage[usenames,dvipsnames]{color}

\newcommand{\beq}{\begin{equation}}
\newcommand{\eeq}{\end{equation}}
\newcommand{\bes}{\begin{subequations}}
\newcommand{\ees}{\end{subequations}}
\newcommand{\bea}{\begin{eqnarray}}
\newcommand{\eea}{\end{eqnarray}}
\newcommand{\ba}{\begin{array}}
\newcommand{\ea}{\end{array}}
\newcommand{\beqn}{\begin{eqnarray*}}
\newcommand{\eeqn}{\end{eqnarray*}}

\newcommand{\om}{\omega}

\newcommand{\nn}{\nonumber}

\newcommand{\tl}[1]{\tilde{#1}}
\newcommand{\bra}{\langle}
\newcommand{\ket}{\rangle}

\def\nn{\nonumber}

\begin{document}

\title{Nonlinearity-induced transition in heat conduction through a topological metamaterial of rotors}
\author{T. R. Vishnu} 
\affiliation{Raman Research Institute, Bangalore 560080, India}
\affiliation{Talent Development Centre, IISc Challakere Campus, Karnataka 577536, India}
\author{Dibyendu Roy}
\affiliation{Raman Research Institute, Bangalore 560080, India}

\begin{abstract}
We investigate heat conduction in a one-dimensional chain of rigid rotors. The rotors are constrained to rotate in a plane about fixed pivot points and coupled by springs, such that in equilibrium, the neighboring rotors lie on opposite sides of the chain axis. The linearized limit of this model valid for small angular displacements, was first introduced by Kane and Lubensky (KL) as a topological mechanical insulator hosting zero-energy vibrational edge modes. We show that the linearized KL chain behaves as a thermal insulator at low temperature in both the topological phases with a finite band gap, and the heat current falls exponentially with the chain length. When the gap vanishes at the topological phase transition, the KL chain becomes a good thermal conductor and conducts heat ballistically. The chain of rotors for arbitrary angular displacements hosts nonlinear solitary waves and distinct topological mechanical phases. Our numerical analysis shows normal (diffusive) heat conduction in all topological phases of the nonlinear chain. Nevertheless, a finite thermal conductivity is achieved for different system sizes in different topological phases of this nonlinear chain.


\end{abstract}

\begin{titlepage}
\maketitle
\end{titlepage}

\section{Introduction}
\label{s:Introduction}
The concept of topological phases and phase transitions in condensed matter physics originated from the work of Berezinskii, Kosterlitz, and Thouless (BKT) in two spatial dimensional classical XY model or rotor (rotator) model almost fifty years ago \cite{thouless1998topological}. The BKT phase transition has been observed experimentally in thin films of superfluid liquid helium \cite{Bishop1978}, thin disordered superconducting granular films \cite{Hebard1980,Epstein1981}, and Josephson junction arrays \cite{Resnick1981}. The XY or rotor model is invariant under a global continuous rotation of all two-dimensional spins or rotors, a continuous global $O(2)$ symmetry. A decade ago, Kane and Lubensky (KL)  \cite{Kane_2013} introduced another topological model of two-dimensional rotors in one spatial dimension (1D) without the continuous global $O(2)$ symmetry. This linearized model is equivalent to the Su-Schrieffer-Heeger (SSH) model for the electronic excitations of polyacetylene that possesses topologically protected electron states at the free ends or interfaces. The KL model consists of $N$ rigid rotors of mass $m$ and length $r$, which are constrained to rotate about fixed pivot points on a 1D lattice with spacing $a$. These rotors are connected by springs with spring constant $k$ so that the alternate rotors in equilibrium make an angle $\bar{\theta}$ and $\pi-\bar{\theta}$ with the normal (see Fig.~\ref{f:KL-model-graphics}). The KL model hosts at least one zero-energy vibrational mode, which can be inferred from the generalized Maxwell's rule. 


For small angular displacement of the rotors from their equilibrium position, we can linearize the change in length of the spring between two rotors in terms of angular displacements of these rotors $(\delta \theta_{n}, \delta \theta_{n+1})$. Such a linearized model's vibrational spectrum  (phonon modes) under periodic boundary conditions (PBC) consists of acoustic and optical branches. The acoustic branch at a small wavevector is gapped except for specific $\bar{\theta}$ values. Moreover, the KL model hosts a zero-energy mode for open boundary conditions (OBC), and the mode is exponentially localized over space at the left or right edge of the rotor chain, depending on $\bar{\theta}$. Thus, the linearized model of KL is a topological mechanical insulator since there are no propagating low-energy phonon modes apart from the exponentially localized zero-energy mode below the gap frequency. Using experimental prototypes of the rotor chain, \textcite{Chen_2014}, however, demonstrated that when the chain was tilted, a zero (soft) mode initially localized at the edge of the chain propagated under the effect of gravity to the opposite end. Such conduction of zero modes under a mechanical perturbation was argued by going beyond the linearized theory of KL, where the carriers are nonlinear solitary wave excitations. \textcite{Chen_2014} further showed the emergence of different topological solitary waves and phases in the nonlinear model of the chain of rotors. These studies have generated huge research interest, including topological mechanics of gyroscopic metamaterials \cite{Nash2015, Wang2015}, non-reciprocal topological solitons in active metamaterials \cite{Veenstra2024,Veenstra2025}, and nonlinearity-induced topological phase transitions \cite{Sone2024,Sone2025}. The primary purpose of this paper is to investigate the nature of heat conduction in the linear and nonlinear chain of rotors under thermal perturbation from the boundaries. 

Heat conduction by phonons in linear and nonlinear low-dimensional lattice models with or without disorder has been extensively explored \cite{Lepri_2003, Dhar_2008}. It is generally believed that 1D momentum-conserving systems manifest an anomalous heat transport (violating Fourier's law of thermal conduction), where the thermal conductivity diverges as a power-law of system size \cite{Narayan_2002}. A normal (diffusive) heat transport can be attained in momentum non-conserving nonlinear models \cite{Hu_1998, Hatano_1999, Tsironis1999, Hu_2000, AOKI2000250, Prosen_2005}. An exception to the above conjecture is the emergence of diffusive transport in the chain of coupled rotors or 1D classical XY chain with translational and continuous global $O(2)$ invariance  \cite{Giardina_2000,Gendelman_2000}, which ensures total (angular) momentum conservation. Nevertheless, the heat conduction by phonons in topological lattices has not been studied much. We thus investigated heat conduction in the chain of rotors hosting topological soft modes and connected to two thermal baths at different temperatures at the edges. We analytically derive the heat current in the KL chain, and it falls exponentially with the chain length at low temperature in the topological phases with a finite band gap. This indicates the KL chain acts as a thermal insulator in the topological phases. When the gap vanishes at the topological phase transition, the KL chain becomes a good thermal conductor and conducts heat ballistically. We further numerically explore classical heat conduction in the nonlinear chain of rotors for arbitrary angular displacement. Our analysis shows diffusive heat conduction in the nonlinear chain. However, we find a finite (system-size independent) thermal conductivity in this nonlinear chain for different chain lengths in different topological phases. Our study reveals a nonlinearity-induced transition from thermal insulator to normal thermal transport in the topological phases of the metamaterial of rotors. The rest of the article is arranged in the following sections and appendices. In Sec .~\ref{TopMat}, we introduce the topological metamaterial of rotors and its topological phases in the linear and nonlinear limit. We discuss heat conduction in the linearized KL chain and nonlinear metamaterial in Sec .~\ref{Heat}. We briefly summarize our findings and comment on future related studies in Sec.~\ref{Sum}. The calculation details of heat conduction in the KL chain is given in App.~\ref{App}. 

\section{Topological metamaterial of rotors}
\label{TopMat}
\begin{figure}[h]
	\begin{center}
	\includegraphics[width=9cm]{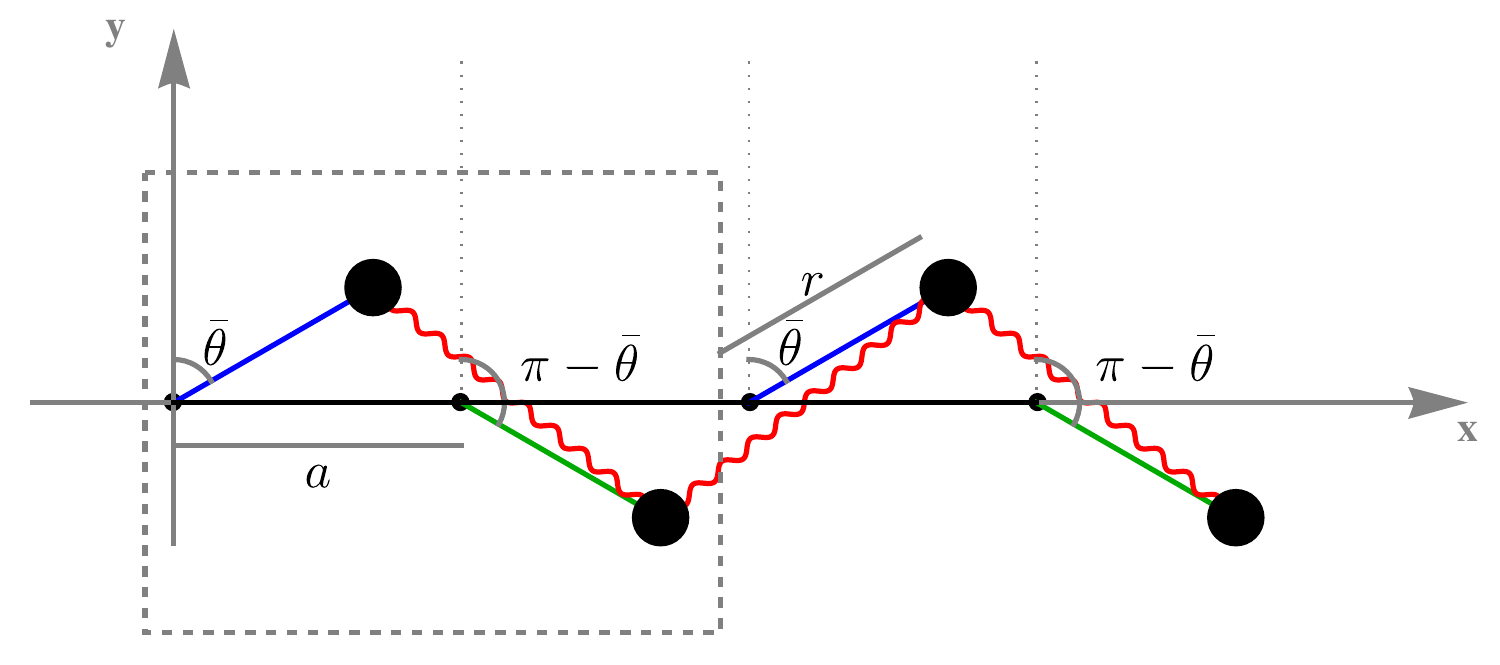}
	\end{center}
	\caption{A cartoon of the topological metamaterial of rotors. The chain is in an equilibrium position, where all rotors have a length $r=0.8$, and their pivot points are separated by a distance $a=1$. Out of the four rotors, the rotors at positions $x=0,2$ (blue) make an angle $\bar{\theta} = \pi/3$ with the $+y$ axis and the rotors at positions $x=1,3$ (green) make angle $ \pi - \bar{\theta} = 2\pi/3$ with the $+y$ axis. 
          The dotted box represents a unit-cell of the lattice. Since this is a right-leaning configuration, we may able to produce nonlinear solitary waves in this chain by perturbing the right-most rotor ($x=3$).}
	\label{f:KL-model-graphics}
	\end{figure}

The Hamiltonian of the chain of $N$ rigid rotors for arbitrary angular displacement $\theta_n$ with the normal (e.g., with the $+y$ axis in Fig.~\ref{f:KL-model-graphics}) and conjugate momentum $p_n$ at site $n$ can be written as
\bea
	H = \sum_{n=1}^{N} \frac{p_n^2}{2m r^2}+\sum_{n=1}^{N-1} \frac{1}{2} k (l_{n,n+1} - \bar{l})^2,\label{Ham1}
	\eea
        where $\bar{l}$ and $l_{n,n+1}$ are, respectively, the equilibrium (unstretched) and stretched length of the spring between two neighbouring rotors at $n$ and $n+1$. We have $\bar{l} = \sqrt{a^2 + 4 r^2 \cos^2(\bar{\theta})}$ for all neighbours when the adjacent rotors make the angles $\theta_n =\bar{\theta}$ and $\theta_{n+1} = \pi -\bar{\theta}$ for odd $n$ (see Fig.~\ref{f:KL-model-graphics}). We can further evaluate
        \bea
	l_{n,n+1}^2 &=& a^2 + 2 r^2 + 2 ar (\sin \theta_{n+1} - \sin \theta_n) \cr
	&&- 2 r^2 \cos(\theta_{n+1} - \theta_{n}).\label{stLength}
	\eea
Assuming the springs being almost rigid, we can rewrite the potential energy term as
	\bea
	\frac{1}{2} k (l_{n,n+1} - \bar{l})^2 &=& \frac{1}{2} \frac{k ((l_{n,n+1} + \bar{l}) (l_{n,n+1} - \bar{l}))^2}{(l_{n,n+1} + \bar{l})^2}  \cr
	&\approx & \frac{\kappa}{2} (l_{n,n+1}^2 - \bar{l}^2)^2, \label{approx}
	\eea
        where, we approximate $(l_{n,n+1} + \bar{l})^2 \approx 4 \bar{l}^2$ and the renormalized spring constant is $\kappa = k/ 4 \bar{l}^2$. Thus, we get the following expression by substituting Eqs.~\ref{stLength} and \ref{approx} in Eq.~\ref{Ham1}:
        \bea
H &=& \sum_{n=1}^{N} \frac{p_n^2}{2m r^2} + \sum_{n=1}^{N-1} 2 \kappa r^4 \bigg[\frac{a}{r} (\sin(\theta_{n+1}) - \sin(\theta_{n}))  \cr
	&&  - \cos(2 \bar{\theta})  - \cos(\theta_{n+1} - \theta_{n}) \bigg]^2.
	\label{Ham2}
        \eea
This Hamiltonian was investigated in \cite{Chen_2014} for explaining the propagation of a soft mode across the chain using a nonlinear conduction mechanism via solitary waves. A linearized version $(H_{KL})$ of the above Hamiltonian for small angular displacement of the rotors from their equilibrium position was examined by KL \cite{Kane_2013}. It can be obtained from Eq.~\ref{Ham2} by substituting $\theta_n = \bar{\theta} - \delta \theta_n, \theta_{n\pm 1} = \pi - \bar{\theta} - \delta \theta_{n \pm 1}$ and taking the linearized limit $\cos(\delta \theta_n) \approx 1$ and $\sin(\delta \theta_n) \approx \delta\theta_n$. 
\bea
H_{KL} &=& \sum_{n =1}^{N} \frac{\delta p_n^2}{2m r^2}  + \frac{k}{2}\sum_{n=1}^{N-1} (v_1 \delta \theta_{n} - v_2 \delta \theta_{n+1})^2, \\
 v_{1,2} &=& \frac{( 2 r \sin(\bar{\theta}) \pm a) r \cos(\bar{\theta})}{\sqrt{a^2 + 4 r^2 \cos^2(\bar{\theta})}},
\eea
where the extension of spring connecting the rotors at site $n$ and $n+1$ is $v_1 \delta \theta_{n} - v_2 \delta \theta_{n+1}$. Due to different equilibrium configuration or angle of rotors at even and odd sites of the chain, the chain can be viewed as a lattice with a basis of two sites. Thus, the vibrational spectrum of the chain consists of two branches, namely acoustic and optical branches, which can be obtained by going to Fourier space for PBC. We find the frequencies $\omega_{\mp}(q_j)$ as 
	\bea
	\omega^2_{\mp}(q_j) = \frac{k}{m r^2} (v_1^2 + v_2^2 \mp 2 v_1 v_2 \cos(q_j a/2)),
	\label{e:dispersion-relation-linear-KL}
	\eea
        where, $q_j = 4 \pi j/ N a$ for integer $j \in (-N/4,N/4]$, with an even numbered of basis. If ${\rm sgn}(v_1) = {\rm sgn}(v_2)$ (this means $2 r \sin(\bar{\theta}) > a$), $\omega_{\mp}(q_j)$ are identified as acoustic (-) and optical (+) branches. However, for ${\rm sgn}(v_1) \neq {\rm sgn}(v_2)$ (e.g., $\bar{\theta} = 0, \pi$ correspond to $v_1 = - v_2$), $\omega_{+}(q_j)$ becomes the acoustic branch. We immediately notice that the acoustic branch is gapped when $v_1 \ne v_2$, and the gap vanishes when $|v_1|= |v_2|$ for $\bar{\theta}=0,\pi$ as $q_j \to 0$. The entire phonon spectrum becomes trivially zero at $\bar{\theta}=\pi/2,3\pi/2$ as $v_1,v_2$ vanish. For small $q_j$, we can express acoustic branch dispersion as $\omega^2_+(q_j) \approx \omega_0^2 + c^2 q_j^2$,  where 
	\bea
\omega_0^2 = \frac{4 kr^2\sin^2(2\bar{\theta})}{m\bar{l}^2},~c^2 = \frac{ka^2(a^2 - 4 r^2 \sin^2(\bar{\theta}))\cos^2(\bar{\theta})}{2 m \bar{l}^2}.\nn\\
	\eea
For OBC, there is a zero energy mode that is localized at the left or right edge of the KL chain depending on whether the rotors are left-leaning $(|v_1| < |v_2|$ for $-\pi < \bar{\theta}<0)$ or right-leaning $(|v_1| > |v_2|$ for $0 < \bar{\theta} < \pi)$. The zero mode corresponding to zero potential energy occurs without the stretching or compression of the spring, i.e., $\delta l_{n,n+1}=0$ within the linear approximation. The zero potential energy mode at the left or right edge of the open rotor chain is related to non-zero or zero winding number of the phase of $q$-space rigidity matrix for PBC when $\bar{\theta} \ne 0,\pm\pi/2,\pi$. These topological features of the KL model highlight its similarity with the SSH model. 


Nevertheless, there are more intriguing features in the mechanical model in comparison to the electrical model, as \textcite{Chen_2014} points out that the infinitesimal zero potential-energy motion related to the localized topological modes at the edges extends to a finite zero-energy motion that transmits through the bulk of the chain. These finite zero-energy motions represent different solitary waves, which emerge in the presence of nonlinearity. \textcite{Chen_2014} have shown three types of solitary waves \cite{Chaikin1995} in the nonlinear metamaterial, which are moving domain walls between two distinct topological mechanical phases consisting of left- or right-leaning rotors. For $2(r/a)\sin \bar{\theta}<1$ in the linkage limit of large spring constant $(k\gg 1)$, the passage of a domain wall flips the direction of rotors from $\theta_n=\bar{\theta}$ to $\theta_n=-\bar{\theta}$ at odd $n$. Such a phase of motion in the metamaterial is the flipper phase. In this phase, the entire chain returns to its initial state when the domain wall has moved the chain back and forth twice. For $2(r/a)\sin \bar{\theta}>2/\sin \bar{\theta}$, the rotors rotate counterclockwise by $\pi$ angle after the passage of a domain wall interpolating between the left- or right-leaning mechanical phases. The emergence of such a domain wall occurs in the spinner phase of the rotor chain. In this case, the rotor chain's initial state is restored after the domain wall has traversed the chain back and forth once. Between these two phases of domain wall motion, there is a special flipper phase, namely a wobbling flipper phase, for $1<2(r/a)\sin \bar{\theta}<2/\sin \bar{\theta}$, when the rotors go past their equilibrium orientations and oscillate after the passage of a domain wall.

\section{Heat conduction}
\label{Heat}
Here, we explore the role of these infinitesimal and finite zero-potential energy motions in heat transport through the chain of rotors. To study heat transport, we connect two Langevin (white noise) heat baths at temperatures $T_L$ and $T_R$, respectively, to the left and right edges of the chain of rotors. It is possible to extract an analytical expression of the heat current in the linearized KL model. We first explore the heat conduction in the linearized model. Later, we numerically explore heat conduction in the nonlinear chain of rotors.

\subsection{Heat conduction in an open KL chain}
\label{s:Heat-conduction-in-an-open-KL-chain}

An open KL chain for $\bar{\theta} \ne 0,\pm\pi/2,\pi$ would behave as a thermal insulator at low temperature $T=(T_L+T_R)/2$ when $k_BT<\hbar \omega_0$, since a localized edge mode can not conduct heat. The heat current then falls exponentially with the chain length $N$. When the gap vanishes for $\bar{\theta} = 0,\pi$, the open KL chain will behave as a good conductor and conduct heat ballistically. Thus, the heat current is independent of chain length. We demonstrate these features of heat conduction in an open KL chain by explicitly calculating the exact heat current. For this, we write the quantum Langevin equations of motion for the rotors' angular displacement as follows: 
	\bea
	m r^2 \delta \ddot{\theta}_1  &=& - k v_1^2 \delta \theta_1 + k v_1 v_2 \delta \theta_2 - \gamma_L \delta \dot{\theta}_1 + \eta_L, \cr
	m r^2 \delta \ddot{\theta}_n &=& -k (v_1^2 + v_2^2) \delta \theta_n  + k v_1 v_2 ( \delta \theta_{n -1} +  \delta \theta_{n+1}), \cr
	m r^2 \delta \ddot{\theta}_N &=& -k v_2^2 \delta \theta_N + k v_1 v_2 \delta \theta_{N-1} - \gamma_R \delta \dot{\theta}_N + \eta_R,
	\eea	 		
        where $n=2,3,\dots,N-1$. Here, $\gamma_{L/R}$ are the coupling strength to the left $(L)$ or right $(R)$ heat baths, and $\eta_{L/R}$ represent the thermal noise from the left $(L)$ or right $(R)$ baths. We define the heat current from the left heat bath into the KL chain as 
	\bea
	J_{\rm KL}(t) =-\delta \dot{\theta}_1(t) (-\gamma_L \delta \dot{\theta}_1(t) + \eta_L(t)).
	\eea	
We can solve the equations of motion of the rotors in the steady-state by Fourier transforming to frequency to rewrite them in a matrix form as $Z(\omega) {\tl \Theta}(\omega) = {\tl \eta}(\omega)$, where ${\tl \Theta}(\omega) = (\delta \tl{\theta}_1(\omega), \delta \tl{\theta}_2(\omega), \cdots, \delta \tl{\theta}_N(\omega))^T$ and $\tl{\eta}(\omega)= (\tl{\eta}_L(\omega), 0, \cdots, 0, \tl{\eta}_R(\omega))^T$. The off-diagonal elements of the tridiagonal matrix $Z(\omega)$ are $-k v_1 v_2$, and $-m r^2 \omega^2 + k(v_1^2 + v_2^2)$ are the diagonal elements of $Z(\omega)$ excluding the first and last elements, which are, respectively, $-m r^2 \omega^2 + k v_1^2 - i \omega \gamma_L$ and $-m r^2 \omega^2 + k v_2^2 - i \omega \gamma_R$.  We define the Fourier transformation as 
	\bea
	\tl{\delta\theta_n}(\omega) = \frac{1}{2\pi}\int_{-\infty}^{\infty} dt \delta\theta_n(t) e^{i \omega t},
	\eea
and similarly for other variables including the noises $\tl{\eta}_{L,R}(\omega)$, which satisfy the following fluctuation-dissipation relations: 
	\bea
	\bra \tl{\eta_b}(\omega) \tl{\eta_{b'}}(\omega') \ket = \frac{ \gamma_b \hbar \omega}{\pi} (1 + f_b) \delta_{b,b'} \delta(\omega	+ \omega'), \label{noise}
	\eea
where, $f_b \equiv f(\omega, T_b) = 1/ (e^{\hbar \omega/ k_B  T_b} - 1)$ with $b=L,R$. 
\begin{figure}[h]       
\includegraphics[width=8cm]{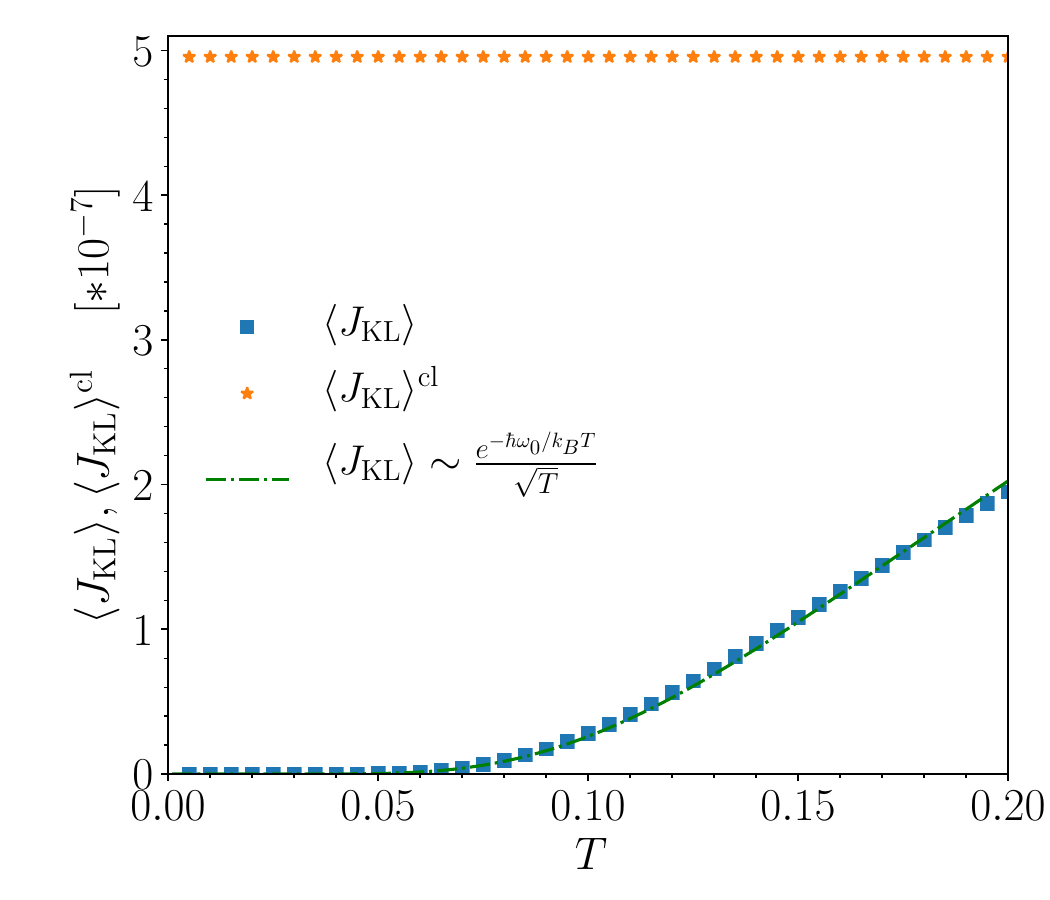}
\caption{Numerically calculated temperature $(T)$ dependence of $\bra J_{\rm KL} \ket$ in Eq.~\ref{HCurrKL1} at low temperatures of an open KL chain of length $N=32$ in the topological phases with a finite band gap ($v_1 \ne v_2$). The dashed-dotted line fit indicates that  $\bra J_{\rm KL} \ket$ scales as $\frac{e^{-\hbar \omega_0/k_B T}}{\sqrt{T}}$ as predicted in Eq.~\ref{Tdep}. The parameters are $m=k=a=1, \gamma_L=\gamma_R=0.2, r=2, \bar{\theta}=\pi/6$. We also show $T$-independent $\bra J_{\rm KL} \ket^{\rm cl}$ from Eq.~\ref{e:classical-linear-KL-current} at the same parameters for a comparison.}
\label{TempSc_QHC_KLgapped}
\end{figure}

\begin{figure}[h]       
 \includegraphics[width=8cm]{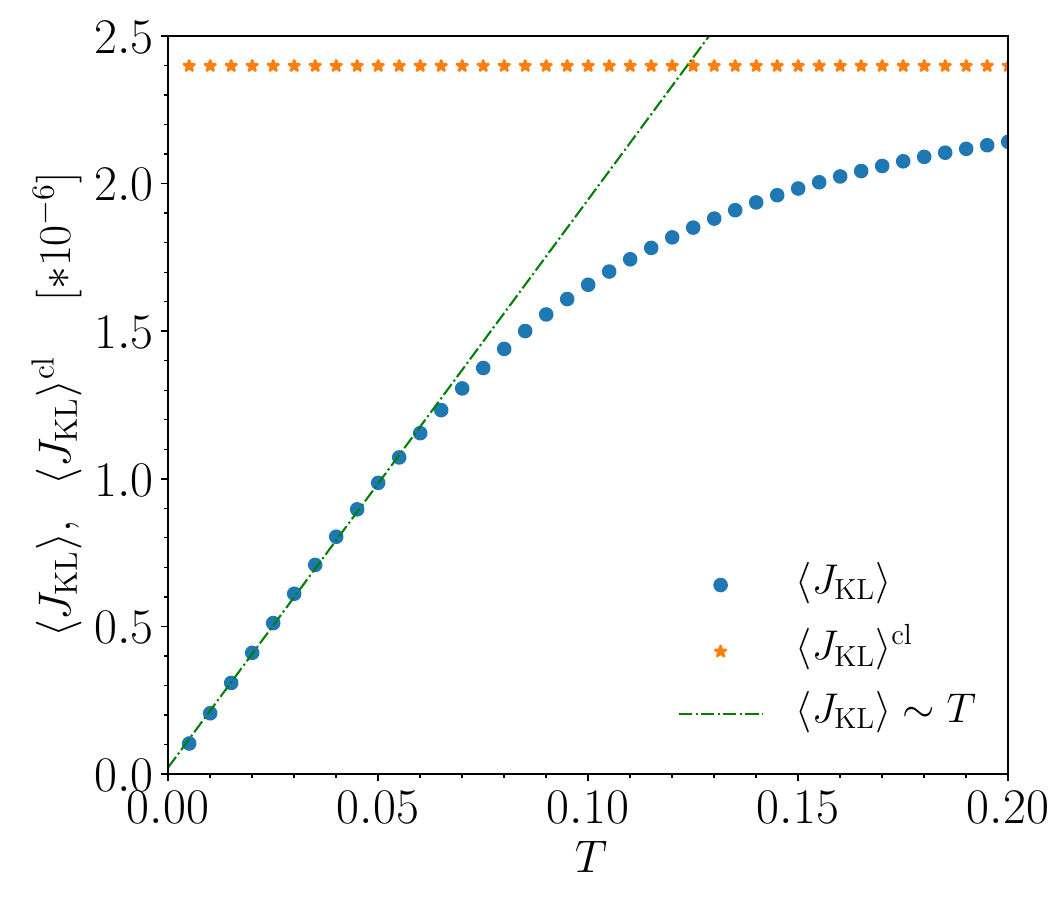}
 \caption{Numerically calculated temperature $(T)$ dependence of $\bra J_{\rm KL} \ket$ in Eq.~\ref{HCurrKL1} at low temperatures of an open KL chain of length $N=32$ at the topological phase transition without a band gap ($|v_1|=|v_2|$). The dashed-dotted line fit depicts that  $\bra J_{\rm KL} \ket$ scales linearly with $T$ at lower temperatures as we predict analytically in App.~\ref{a:Temperature-dependence-of-quantum-current}. The parameters are $m=k=a=1, \gamma_L=\gamma_R=0.2, r=2, \bar{\theta}=0$. $\bra J_{\rm KL} \ket$ approaches $T$-independent $\bra J_{\rm KL} \ket^{\rm cl}$ from Eq.~\ref{e:classical-linear-KL-current} for the same parameters at relatively higher temperatures.}
\label{TempScal_QHC_KLgapless}
\end{figure}

         Thus, we get the formal solution ${\tl \Theta}(\omega) = G(\omega) \tilde{\eta}(\omega)$, where $G(\omega)=Z^{-1}(\omega)$ is a retarded Green's function. 
Using the formal solution, we find the steady-state heat current after taking an average over the noises from the baths as
	\bea
	\bra J_{\rm KL} \ket = \int_{-\infty}^{\infty} d\omega |[G(\omega)]_{1,N}|^2 \gamma_L \gamma_R  \frac{\hbar \omega^3}{\pi} (f_L - f_R),
	\eea 
        where $\langle \dots \rangle$ denotes an averaging over the noises. In the linear response regime when $\Delta T \ll T$ with $\Delta T=T_L-T_R$, we can simplify $\bra J_{\rm KL} \ket$ as (assuming $\gamma_L = \gamma_R = \gamma$)
\bea
\bra J_{\rm KL} \ket &=& \frac{k_B \Delta T \gamma^2}{\pi} \int_{-\infty}^{\infty} d\omega \bigg[ |[G(\omega)]_{1,N}|^2 \left( \frac{\hbar \omega^2}{2 k_B T}\right)^2 \nn\\&&{\rm cosech}^2 \left( \frac{\hbar \omega}{2 k_B T}\right) \bigg].\label{HCurrKL1}	
\eea
An expression of the retarded Green's function to evaluate this integral can be obtained by applying the methods in Refs.~\cite{Dhar_2006, Hu_1996} for the matrix $Z(\omega)$ (see App.~\ref{a:Classical-Heat-current-in-Linarized-KL-model}). We then take a large $N$ limit \cite{Roy_2008, Vishnu_2024} to find the leading-order $T$-dependence of $\bra J_{\rm KL} \ket$ at low temperatures for the gapped acoustic branch spectrum, e.g., $v_1 \ne v_2$ as (see App.~\ref{a:Temperature-dependence-of-quantum-current})
	\bea
	\bra J_{\rm KL} \ket \sim \frac{e^{-\hbar \omega_0/k_B T}}{\sqrt{T}}, \quad \omega_0 = \sqrt{\frac{k}{mr^2}}(v_1 - v_2).\label{Tdep}
	\eea
In Fig.~\ref{TempSc_QHC_KLgapped}, we numerically verify our analytical prediction of $T$-dependence of $\bra J_{\rm KL} \ket$ in Eq.~\ref{Tdep}. We observe from Eq.~\ref{Tdep} that the linearized KL chain behaves as an insulator in the linkage limit of rigid springs ($k \to \infty$) at higher temperatures too (even though edge modes are present, they are highly localized). There can be heat conduction in this limit only by going beyond the linear model  as explained by \textcite{Chen_2014} for conduction under mechanical perturbation. The linearized KL chain becomes a good conductor even at low temperatures when the acoustic branch becomes gapless, and $\bra J_{\rm KL} \ket$ then depends linearly on temperature as shown in Fig.~\ref{TempScal_QHC_KLgapless}. The data that support the findings of this article are available at \cite{Data}. 

\begin{figure}[t]
	\begin{center}
	\includegraphics[width=8cm]{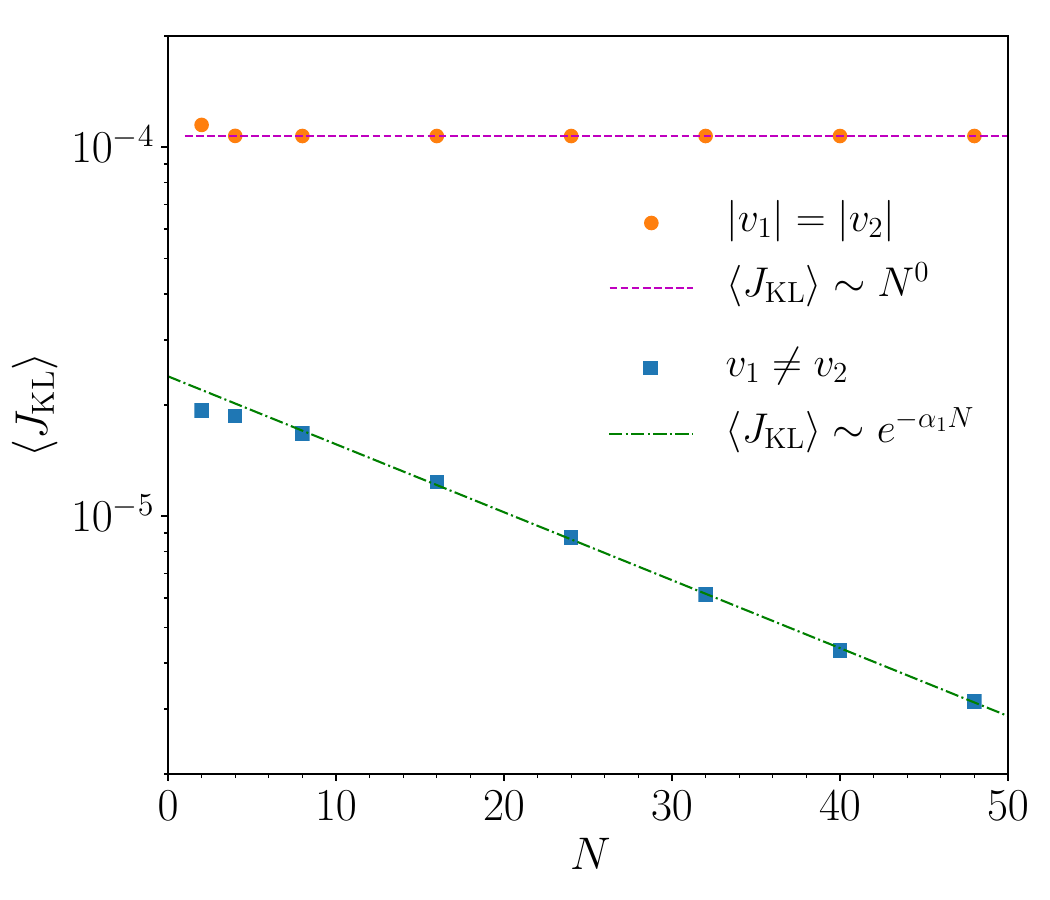}
	\end{center}
	\caption{Log-linear plot of system-size $(N)$ scaling of $\bra J_{\rm KL} \ket$ in Eq.~\ref{HCurrKL1} at low temperatures of an open KL chain in the topological phases with a finite band gap ($v_1 \ne v_2$) and at the topological phase transition without gap ($|v_1|=|v_2|$). The fits (dashed or dashed-dotted lines) indicate that  $\bra J_{\rm KL} \ket$ falls exponentially with $N$ in the topological phases and $\bra J_{\rm KL} \ket$ is $N$-independent at the topological phase transition. The parameters are $m=k=1, \gamma_L=\gamma_R=0.2$ for both the plots, and $r=1,a=0.025, \bar{\theta}=\pi/3, T=0.0075, \Delta T=0.005$ for the squares and $r=2, a=1, \bar{\theta}=0, T=0.055, \Delta T=0.01$ for the circles.}
	\label{LengthScal_QHC_KL}
	\end{figure}

We can also find the $N$-dependence of $\bra J_{\rm KL} \ket$ at low temperatures from Eq.~\ref{HCurrKL1} by finding the $N$-dependence of $|[G(\omega)]_{1,N}|^2$ at small frequencies (see App.~\ref{a:System-size-dependence-of-quantum-current}). We numerically observe that for temperatures below $\hbar \omega_0$ (i.e., $k_B T<\hbar \omega_0$), $\bra J_{\rm KL} \ket$ is exponentially falling with increasing $N$ for $v_1 \neq v_2$. For a relatively large $N$ and a small wavevector $q$, we can express the Green's function as
	\bea
	|G_{1,N}|^2 \approx \frac{m^2 r^4 v_1^2v_2^2 \sin^2(a{\tl q})}{((|{\tl c} + i {\tl d}({\tl q})|)\sin(N a {\tl q}))^2} \sim e^{ \mp 2i(N-1){\tl q}},
	\label{e:N-dependence-of-G1N}
	\eea	
where $2 \cos(a {\tl q})= -(m r^2 \omega^2/k v_1 v_2) + (v_1/v_2 + v_2/v_1), {\tl c}=-(k m r^2v_1v_2 + \gamma^2)(v_1  - v_2)^2, {\tl d}(q)= (v_1 - v_2)^2\gamma \sqrt{k m r^2 (v_1^2+v_2^2 - 2v_1v_2 \cos (a {\tl q}))}$. Then, we substitute ${\tl q} = \mp i(|v_1-v_2|)/(a\sqrt{v_1v_2})$ for the zero energy modes of the open KL chain to find $|G_{1,N}|^2 \sim e^{-2(N-1)|v_1-v_2|/a\sqrt{v_1v_2}}$. Plugging this form of $|G_{1,N}|^2$ in Eq.~\ref{HCurrKL1}, we find $\bra J_{\rm KL} \ket$ at low temperatures is decreasing exponentially with $N$ when $v_1 \neq v_2$. This further indicates thermal insulating nature of the linear KL chain in the topological phases. We also notice that the quantum current of linear KL model is independent of the system size if $|v_1| = |v_2|$  (see App.~\ref{a:System-size-dependence-of-quantum-current} for details) indicating ballistic transport at low temperatures. In Fig.~\ref{LengthScal_QHC_KL}, we have further numerically checked $N$-scaling of $\bra J_{\rm KL} \ket$ in Eq.~\ref{HCurrKL1} both in the gapped topological phases and at the gapless topological phase transition. These $N$-scalings obtained from numerics in Fig.~\ref{LengthScal_QHC_KL} agree well with our above analytical predictions. 

An ordered harmonic chain without spectral gap generally shows ballistic heat conduction (or thermal conductivity linearly diverging with $N$), which can be argued by the stability of linear Fourier modes and the absence of coupling between conducting modes. No thermal gradient is formed across the system in such a case. We prove rigorously in the App.~\ref{a:Classical-Heat-current-in-Linarized-KL-model} that the linear KL chain shows ballistic heat transport at high temperatures $(k_B T\gg \hbar \omega_0)$ both in the topological phases and the phase transition point. The linear response heat current in Eq.~\ref{HCurrKL1} in the high temperature classical limit reads as
	\bea
	\bra J_{\rm KL} \ket^{\rm cl} = \frac{k_B \Delta T \gamma^2}{\pi} \int_{-\infty}^{\infty} d\omega \omega^2 |[G(\om)]_{1,N}|^2,
	\label{e:classical-linear-KL-current}
	\eea
        where, we set $ (\hbar \omega/2 k_B T)^2{\rm cosech}^2 (\hbar \omega/2 k_B T) \to 1$. Applying the expression of the retarded Green's function in App.~\ref{a:Classical-Heat-current-in-Linarized-KL-model} at large $N$ limit, we find an analytical expression of $\bra J_{\rm KL} \ket^{\rm cl}$ as 
        \begin{widetext}
	\bea
	\bra J_{\rm KL} \ket^{\rm cl} = \frac{k_B \Delta T k \Gamma ^2 \left(2 \gamma ^2 \Omega ^2+2 \Gamma ^2 \left(2 \gamma ^2-k m r^2 \Omega \right) \left(\sqrt{1-\frac{4 \gamma ^4 \Omega ^2}{\Gamma ^2 \left(k m r^2 \Omega -2 \gamma ^2\right)^2}}-1\right)- 2 \gamma ^2 \Omega  \sqrt{\Omega ^2-4 \Gamma ^2}\right)}{4 \gamma \Omega \left(\gamma ^2   \Omega ^2 +  \Gamma ^2   \left(k m r^2 \Omega -2 \gamma ^2\right)\right)},
	\label{e:classical-current-KL}
	\eea\end{widetext}
where, $\Omega = v_1^2 + v_2^2$ and $\Gamma = v_1 v_2$. 

\subsection{Heat conduction in the nonlinear metamaterial}
\label{s:Heat-conduction-Nolinear}
For the nonlinear metamaterial, \textcite{Chen_2014} have argued that the flipper and spinner solitary waves can be described by solutions to the $\phi^4$ and sine-Gordon equations, respectively, which emerge in the continuum limit of the constraint equation $\delta l_{n,n+1}=0$, when going beyond the linear approximation. Therefore, the nonlinear topological metamaterial is integrable (in the spinner phase), hosting topological solitary waves in the continuum limit. It is imperative to point out that the shape of the solitary waves of the sine-Gordon equation remains invariant under collisions. These special solitary waves are known as solitons. The solitary waves of the $\phi^4$ equation are not solitons \cite{Chaikin1995}. The integrable models are expected to show ballistic thermal transport due to the presence of nonlinear stable modes (solitons) and complete integrability \cite{Prosen_2005}. \textcite{Prosen_2005} showed ballistic heat transport in the spatially discrete version of the sine-Gordon model by Izergin and Korepin, which preserves the integrability of the continuum limit, and it has an on-site potential breaking total momentum conservation. Thus, nonlinear models, even without total momentum conservation, can show non-diffusive heat transport. The existence of an inﬁnite number of independent conserved quantities ensures ballistic transport \cite{van_Beijeren_2012}. 

	\begin{figure}[h]
	\begin{center}
        \includegraphics[width=9cm]{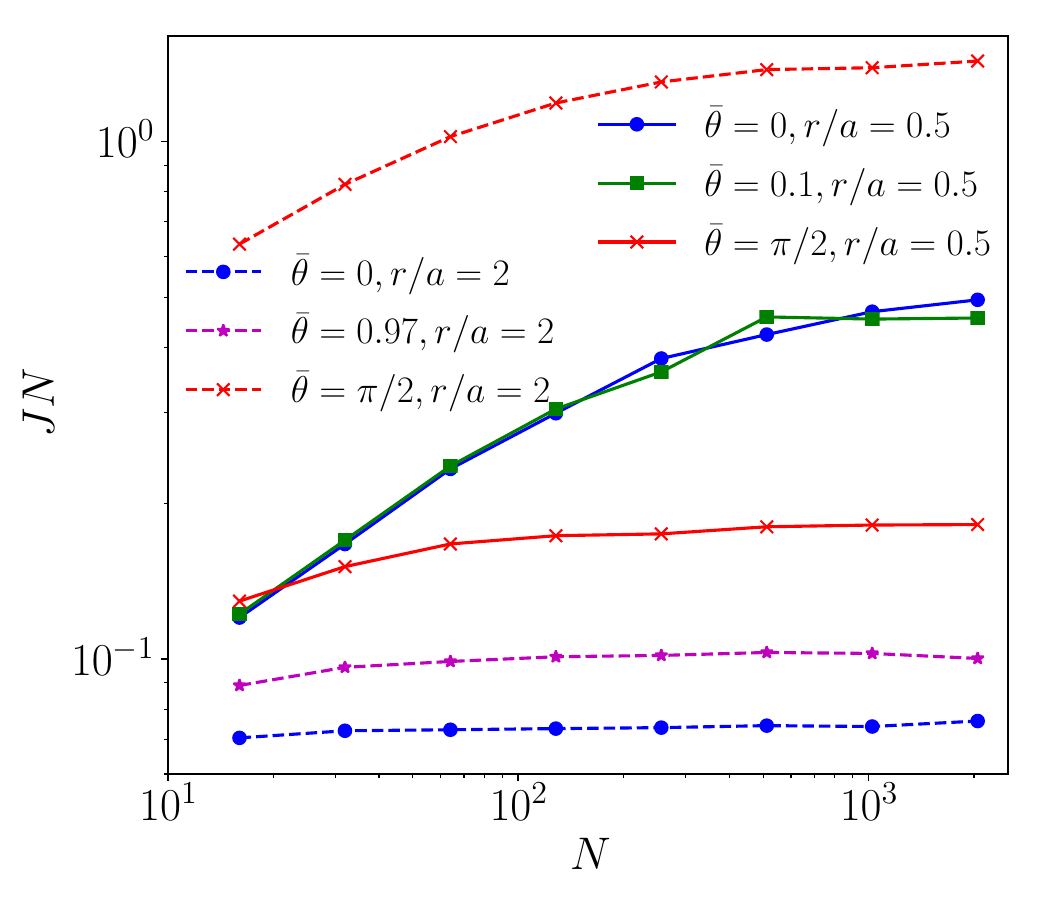}
	\end{center}
	\caption{System-size $(N)$ scaling of $JN$ in a nonlinear metamaterial of rotors for different values of equilibrium angle $\bar{\theta}$ of the rotors and $r/a$, where $J$ is steady-state heat current. Other parameters are $m=k=1, \gamma_L=\gamma_R=0.5, T_L=0.01,T_R=0.5$. $J$ shows diffusive scaling $(J \sim N^{-1})$ at longer $N$ for $\bar{\theta}=0, 0.97, \pi/2$ when $r/a=2$ and $\bar{\theta}=\pi/2, 0.1$ when $r/a=0.5$. For $\bar{\theta}=0$ when $r/a=0.5$, $J$ scales superdiffusively within our finite simulated lengths, and the scaling is $J \sim N^{-0.85}$.}
	\label{NLKL_Heat}
	\end{figure}

        The Frenkel–Kontorova model is a nonintegrable (chaotic), discrete version of the sine-Gordon model, and it consists of harmonic nearest-neighbor interactions and a periodic on-site substrate potential, implying the absence of translational invariance or momentum non-conservation. The Frenkel–Kontorova model and the discrete $\phi^4$ model show diffusive thermal conduction (or a system-size independent thermal conductivity) \cite{Hu_1998, Hu_2000, AOKI2000250}, which has been used for a general argument for diffusive heat conduction in momentum non-conserving models. The nonlinear metamaterial in Eq.~\ref{Ham2} in its discrete form is nonintegrable and chaotic, which is indicated by $N-1$ positive Lyapunov exponents in different parameter regimes, including the spinner and flipper phases. In fact, positive Lyapunov exponents lead small differences in the initial states to grow exponentially in time, resulting in the development of chaos. Nonintegrable models have shown diffusive and anomalous (superdiffusive) thermal transport depending on the  properties of the models and temperature or other parameters \cite{Casati_1984,Prosen_1992, Lepri_1997, Lepri_1998-1, Lepri_1998-2, Hu_1998, Hatano_1999, Dhar_2001-1, Savin_2002, Grassberger_2002, Casati_2003, Prosen_2005, van_Beijeren_2012, DRoy_2012}.
        
We investigate classical heat transport in the nonlinear metamaterial connected to two Langevin heat baths at different temperatures. For this, we numerically evaluate the time-evolution of $\theta_n(t)$ using the following equations of motion for $H$ in Eq.~\ref{Ham2} from some arbitrary initial conditions:
\bea
m r^2 \ddot{\theta}_1  &=& - \frac{\partial H}{\partial \theta_1} - \gamma_L \dot{\theta}_1 + \eta_L, \label{ne1} \\
	m r^2 \ddot{\theta}_n &=& - \frac{\partial H}{\partial \theta_n},~~ n=2,3,\dots,N-1, \label{ne2}\\
	m r^2 \ddot{\theta}_N &=& - \frac{\partial H}{\partial \theta_N} - \gamma_R \dot{\theta}_N + \eta_R, \label{ne3}
	\eea	 		
        where $\gamma_{L/R}$ are again the coupling strength to the left $(L)$ or right $(R)$ heat baths, and $\eta_{L/R}$ denote the thermal white noise from the left $(L)$ or right $(R)$ baths, which satisfy the relations in Eq.~\ref{noise} in the high temperature classical limit. The  nonlinear chain reaches a steady state after a transient process over many time steps, which depends on the chain length (e.g., $10^7$ steps for $N \sim 100$). We calculate the local heat current at the bond between neighbouring rotors at $n,n+1$ using the formula
        \bea
J_{n}(t)=-\dot{\theta}_n\frac{\partial V(\theta_n,\theta_{n+1})}{\partial \theta_n},
        \eea
where $ V(\theta_n,\theta_{n+1})$ denotes the potential of $H$ in Eq.~\ref{Ham2} when rewritten as $H = \sum_{n=1}^{N} \frac{p_n^2}{2m r^2} + \sum_{n=1}^{N-1} V(\theta_n,\theta_{n+1})$. In the steady state of the system, the long-time averaged $\bar{J_{n}}$ is independent of $n$. We take mean of these currents over the full system to find $J=\sum_{n=1}^{N-1}\bar{J_{n}}/(N-1)$. The local temperature of $n^{\rm th}$ rotor can be found from $T_n=\overline{p_n^2}/(mr^2)$, where the over-line denotes the long-time average. The Markovian nature of the Langevin heat baths with white noises in our classical thermal transport analysis by numerical simulation leads to significant differences compared with the quantum transport analysis using non-Markovian baths with colored noises for the KL model. Therefore, we can not recover the quantum thermal transport behavior in the linearized KL model from the classical simulations of the nonlinear rotor chain in the limit of $T\to 0$. 

Since the discrete nonlinear metamaterial is nonintegrable and total (angular) momentum non-conserving, we expect it to show diffusive thermal transport or $J \sim N^{-1}$. We investigate system-size $(N)$-dependence of the steady-state heat current $J$ in different parameter regimes representing the topological spinner and flipper phase, and the nontopological case of $\bar{\theta}=0$, which corresponds to a gapless acoustic dispersion in the linear KL model. We also consider $\bar{\theta}=\pi/2$, where the entire phonon spectrum of the linear KL model vanishes. Our main findings are shown in Fig.~\ref{NLKL_Heat}, where we plot $JN$ vs. $N$. It clearly shows $JN$ is a constant for larger $N$ in the spinner phase (e.g., $\bar{\theta}=0.97$ when $r/a=2$), indicating diffusive scaling of $J$  as a function of $N$. This is also the case for $\bar{\theta}=0$ when $r/a=2$ and $\bar{\theta}=\pi/2$ when $r/a=0.5$. We clearly notice diffusive heat transport in our nonequilibrium simulation, manifested by a constant $JN$ vs. $N$ for larger $N$ values. Further, $JN$ nearly becomes a constant within our simulated lengths for $\bar{\theta}=\pi/2$ when $r/a=2$ and in the flipper regime (e.g., $\bar{\theta}=0.1$ when $r/a=0.5$). Within our simulated lengths,  $JN$ is not a clear constant for $\bar{\theta}=0$ when $r/a=0.5$. Nevertheless, we expect $J$ to reach diffusive scaling even for these parameters over longer lengths. We further observe in Fig.~\ref{NLKL_Heat} that thermal conductivity is higher for a higher $\bar{\theta}$ when $r/a=2$, and the trend is reversed for $r/a=0.5$. One exciting feature of $JN$ vs. $N$ plots in Fig.~\ref{NLKL_Heat} is different length scales, when the diffusive scaling of $J$ emerges in different parameter regimes. We try to understand it by following a small-angle expansion of $H$ around the equilibrium angle $\bar{\theta}$.

We expand $H$ in Eq.~\ref{Ham2} in orders of $\delta \theta_n$ and $\delta \theta_{n+1}$ using $\theta_{n} = \bar{\theta} - \delta \theta_n$, $\theta_{n+1} = \pi - \bar{\theta} - \delta\theta_{n+1}$. Thus, $H$ simplifies around $\bar{\theta}=0$ as
\bea
H &=& \sum_{n=1}^{N} \frac{\delta p_n^2}{2 m r^2} + \sum_{n=1}^{N-1} 2 \kappa r^4 \bigg[ \frac{a}{r} \bigg( \delta \theta_n + \delta \theta_{n+1} - \frac{\delta \theta_n^3}{3!} \nn\\ &-& \frac{\delta \theta_{n+1}^3}{3!} \bigg)- \frac{(\delta \theta_{n} - \delta \theta_{n+1})^2}{2} \bigg]^2. \label{Exp0}
\eea
We observe significantly different behavior of $JN$ with $N$ in Fig.~\ref{NLKL_Heat} for $r/a=0.5,2$ when $\bar{\theta}=0$. The coefficients of three terms of potential $V$ in Eq.~\ref{Exp0} are significant in determining the thermal transport. These terms are $(\delta \theta_n + \delta \theta_{n+1})^2, (\delta \theta_n + \delta \theta_{n+1})(\delta \theta_{n} - \delta \theta_{n+1})^2$ and $(\delta \theta_{n} - \delta \theta_{n+1})^4$. The first term forms the potential of the KL model for $\bar{\theta}=0$, leading to a gapless acoustic spectrum and ballistic thermal transport in the KL model. For a small $r/a$ (e.g., $r/a=0.5$ or $a/r=2$), the coefficient of the first term is larger than the others. Thus, a longer length-scale is required for the emergence of a constant $JN$ vs. $N$ by relatively weaker nonlinear terms. On the contrary, the coefficient of the second term is larger than the others for $r/a=2$ or $a/r=0.5$, which leads to a shorter length-scale for the emergence of a constant $JN$ vs. $N$ by stronger nonlinear terms.


Next, we consider $\bar{\theta} = \pi/2$, where we keep only upto the second order terms in the series expansion of trigonometric functions, e.g., $\sin$ and $\cos$. Thus, we get 
	\bea
	H &=& \sum_{n=1}^{N} \frac{\delta p_n^2}{2 m r^2} + \sum_{n=1}^{N-1} 2 \kappa r^4 \bigg[ \frac{a}{r} \bigg( \frac{\delta \theta_n^2}{2} - \frac{\delta \theta_{n+1}^2}{2} \bigg) \nn\\&+& \frac{(\delta \theta_{n} - \delta \theta_{n+1})^2}{2} \bigg]^2. \label{ExpP}
	\eea
        It is immediate from Eq.~\ref{ExpP} that the first term $(a/r)^2(\delta \theta_n^2/2 - \delta \theta_{n+1}^2/2)^2$ in the potential is dominating for smaller $r/a$, and this term breaks translation symmetry of the model. On the other hand, the translation-invariant second term in the potential in Eq.~\ref{ExpP} dominates for larger $r/a$. Therefore, the nonlinear metamaterial with $\bar{\theta} = \pi/2$ is expected to show diffusive behavior within a shorter length for $r/a=0.5$ compared to $r/a=2$.

\begin{figure}[h]
\begin{center}
\includegraphics[width=8cm]{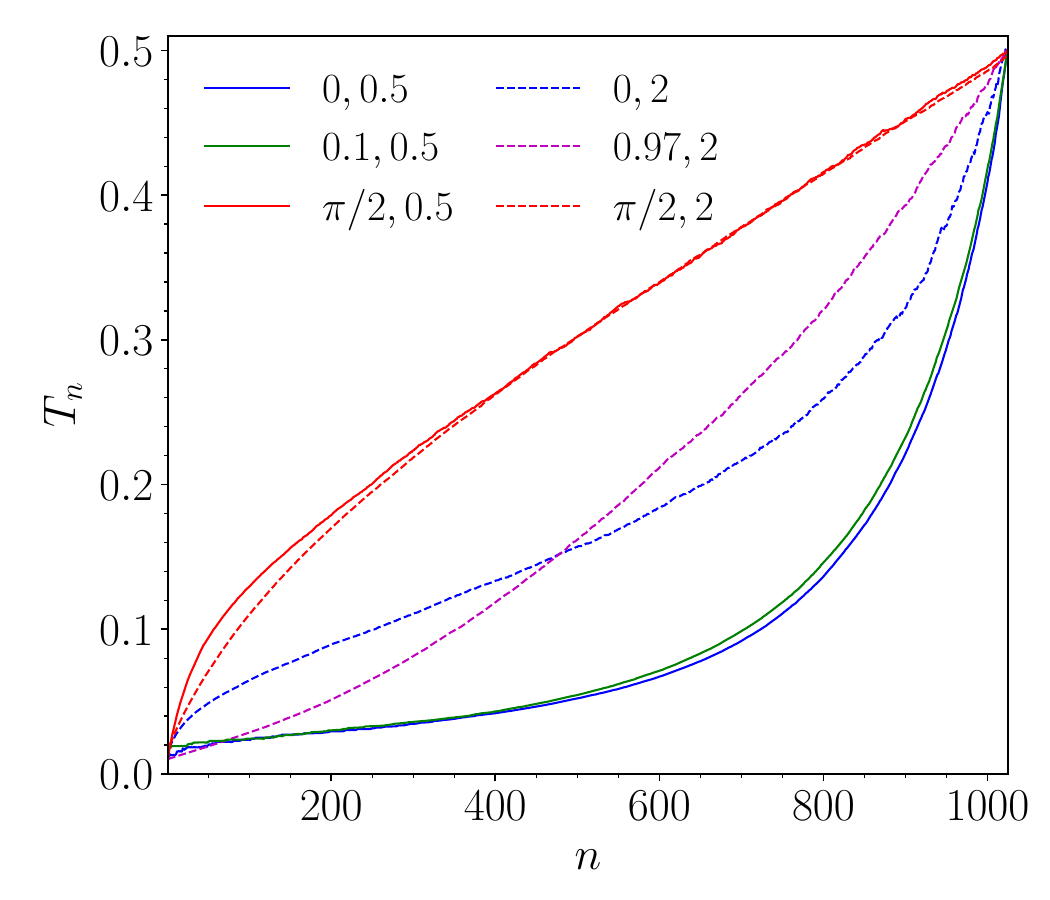}
\end{center}
\caption{The profile of steady-state local temperature $T_n$ of $n^{\rm th}$ rotor of a nonlinear metamaterial of rotors of length $N=1024$. The six lines are for different equilibrium angle $\bar{\theta}$ and $r/a$ as given in the figure. Other parameters are $m=k=1, \gamma_L=\gamma_R=0.5, T_L=0.01,T_R=0.5$. The long-time average is taken over $7\times 10^7$ steps after the transient time of $7 \times 10^7$ steps.}
\label{NLKL_Temp}
\end{figure}

The profile of steady-state local temperature $T_n$ of the $n^{\rm th}$ rotor of a nonlinear metamaterial of rotors of length $N=1024$  is shown in Fig.~\ref{NLKL_Temp}. While we do not find a clear linear $T_n$ with $n$ in any of these six parameter regimes discussed in Fig.~\ref{NLKL_Heat}, the temperature profile for $\bar{\theta}=0,0.1$ when $r/a=0.5$ shows the most deviation from the linear profile, which gives some justification for superdiffusive scaling of $J$ within our finite simulated lengths.          

\section{Summary and outlook}
\label{Sum}
In summary, our findings on heat transport in linear and nonlinear metamaterials of rotors are in accordance with the primary features of these metamaterials under mechanical perturbation, namely, a nonlinearity induced transition from insulation to conduction \cite{Kane_2013, Chen_2014}.  \textcite{Chen_2014} introduced a classification of topological phases of the nonlinear metamaterials based on the nature of solitary waves. We explicitly calculated the heat current and its scaling with system size in these linear and nonlinear models under thermal perturbation from the boundaries. The heat current in the KL chain decays exponentially with length in the topological phases and is independent of length at the topological phase transition.  Thus, the heat current in the linear KL chain can naturally characterize the topological phases and phase transitions in the KL chain, even though it can not differentiate the two topological phases. On the contrary,  the classical thermal transport in nonlinear long chains is not sensitive to the topological phases and phase transitions. We find that the heat current falls linearly with length in different topological phases of  \textcite{Chen_2014}. We have not observed any significant difference in the system-size scaling of heat current across these phase boundaries. Even low-temperature classical transport in the nonlinear chain is diffusive, even though it may take a long length scale to achieve such scaling. Therefore, classical thermal transport measurement can not be applied to separate different topological phases of interacting or nonlinear metamaterials. In fact, it is interesting to find out which physical quantity can be used to detect different topological phases of  \textcite{Chen_2014}. Studying quantum heat transport in the nonlinear chain is an exciting challenge in this context. The Lyapunov exponents are zero for the KL chain and mostly non-zero in the nonlinear chain. Analyzing thermal transport in two-dimensional systems with anomalous Hall effect behavior or non-reciprocal breathers \cite{Veenstra2024,Veenstra2025} would also be helpful.

\vspace{0.5cm}

\appendix

\section{Heat current in an open KL chain}\label{App}
In this appendix, we derive the temperature and system-size dependence of the quantum heat current of the open KL chain, along with the analytical expression for its high-temperature classical heat current.

\subsection{Temperature dependence of quantum heat current}
\label{a:Temperature-dependence-of-quantum-current}

The formal expression of quantum heat current in the linear response regime in Eq.~\ref{HCurrKL1} can be expressed in terms of system parameters as (here $\Omega = v_1^2 + v_2^2$ and $\Gamma = v_1 v_2$)
\begin{widetext}
	\bea
	\bra J_{\rm KL} \ket = \frac{\hbar^2 \Delta T}{2 \pi k_B T^2} \int_{0}^{\pi/a} d{\tl q} \frac{a \gamma k^2 \Gamma^3 \sin^2(a {\tl q})}{m r^2 (\Gamma (k m r^2 \Omega - 2 \gamma^2) + 2 \Omega \gamma^2 \cos(a {\tl q}))} {\rm cosech}^2  \left( \frac{\hbar \omega}{2 k_B T}\right). \label{QHC1}	
	\eea
\end{widetext}
Here, we re-express $|G_{1,N}(\omega)|^2$ in terms of the variable $\tl{q}$, which is explained in detail in App.~\ref{a:Classical-Heat-current-in-Linarized-KL-model}. In the small $\tl{q}$ limit, we have
	\bea
	\omega^2(\tl{q}) &=& \frac{k}{mr^2} (v_1^2 + v_2^2 - 2 v_1 v_2 \cos(a \tl{q})) \cr 
	&\approx& \frac{k(v_1 - v_2)^2}{mr^2}  + \frac{k v_1 v_2 a^2}{m r^2}  \tl{q}^2 = \omega_0^2 + \lambda \tl{q}^2.
	\eea
Thus, we re-express Eq.~\ref{QHC1} in this limit of small  $\tl{q}$ as 
\begin{widetext}
	\bea
	\bra J_{\rm KL} \ket \approx \frac{2 \hbar^2 \Delta T}{\pi k_B T^2} \int_{0}^{\pi/a} d{\tl q} \frac{a^3 \gamma k^2  \Gamma^3 {\tl q}^2 e^{-\hbar \omega_0/k_B T} e^{-\hbar \lambda {\tl q}^2/2 k_B T \omega_0}}{m r^2(\Gamma (k m r^2 \Omega - 2 \gamma^2) + 2 \Omega \gamma^2 \cos(a {\tl q}))}.\label{QHC2}
	\eea
\end{widetext}
 Here, we used the approximation ${\rm cosech}^2(\hbar \omega/k_B T) \approx 4 e^{\hbar \omega_0/k_B T} e^{-\hbar \lambda {\tl q}^2/ 2 k_B T \omega_0}$ to get Eq.~\ref{QHC2} from Eq.~\ref{QHC1}. We then substitute 
	\bea
	x = \frac{\hbar}{k_B T} \sqrt{\frac{k}{mr^2}} \frac{v_1 v_2 a^2 {\tl q}^2}{(v_1 - v_2)},
	\eea
in Eq.~\ref{QHC2} to obtain the leading-order temperature dependence of the quantum heat current, which is given in Eq.~\ref{Tdep}, for the gapped case $v_1 \ne v_2$.

Similarly, for the gapless case ($|v_1| = |v_2|$), the temperature dependence of $\bra J_{\rm KL} \ket$ can be obtained as discussed below. We notice in this case: 
	\bea
	\omega^2 = \frac{k}{m r^2} (2 v_1^2 ( 1 - \cos({\tl q} a))) =  \frac{4 v_1^2 k}{m r^2} \sin^2\left(\frac{a {\tl q} }{2}\right).
	\eea   
Thus, a similar analysis like above gives us 
\begin{widetext}
	\bea
	\bra J_{\rm KL} \ket = \frac{\hbar^2 \Delta T}{2 \pi k_B T^2} \int_{0}^{\pi/a} d{\tl q} \frac{a \gamma k^2 \Gamma^2 \sin^2(a {\tl q})}{2 mr^2(k m r^2 \Gamma -  \gamma^2 + 2 \gamma^2 \cos(a {\tl q}))} {\rm cosech}^2  \left( \frac{\hbar \omega}{2 k_B T}\right).	
	\eea
By taking the small ${\tl q}$ limit, we get
	\bea
	\bra J_{\rm KL} \ket \approx \frac{\hbar^2 \Delta T}{4 \pi k_B T^2} \int_{0}^{\pi/a} d{\tl q} \frac{a \gamma k^2 \Gamma^2 a^2 {\tl q}^2}{(k m r^2 \Gamma -  \gamma^2 + 2 \gamma^2 - \gamma^2 a^2 {\tl q}^2)} {\rm cosech}^2  \left( \frac{\hbar \omega}{2 k_B T}\right).\label{QHC3}	
	\eea
\end{widetext}
We next substitute $x = (\hbar \omega)/2 k_B T$ in Eq.~\ref{QHC3} to find the temperature dependence as $\bra J_{\rm KL} \ket \sim T$ when $v_1 = v_2$. To obtain this result, we make the assumption $k m r^2 \Gamma - \gamma^2 > 0$, which is satisfied mostly for small coupling $\gamma$ compared to the system parameters $k, m, r$ and $\Gamma$.

\subsection{System-size dependence of quantum heat current}
\label{a:System-size-dependence-of-quantum-current}

In order to get the $N$-dependence of $\bra J_{\rm KL} \ket$ for $v_1 \neq v_2$, we need to find the $N$-dependence of $|[G(\omega)]_{1,N}|^2$ at small frequencies. We numerically observe that $\bra J_{\rm KL} \ket$ is exponentially falling with $N$ for temperatures below $\hbar \omega_0$, i.e., $k_B T < \hbar \omega_0$. To derive this result analytically for the KL chain, we find the following expression for the Green's function when $\gamma_L = \gamma_R =\gamma$:
\begin{widetext} 
\footnotesize
	\bea
	|G_{1,N}({\tl q})|^2 &=& \frac{\Gamma^2 m^2 r^4 \sin^2(a {\tl q})}{\bigg( \bigg| \left(k m r^2 \Gamma^2 - \gamma^2 \Omega \right) \sin((N-1)a {\tl q}) + \Gamma \left(\gamma^2 - k m r^2 \Omega \right) \sin(N a {\tl q}) - 2  \Gamma  \sin(N a {\tl q}) )} \cr
	&& {+ \Gamma^2 k m r^2 \sin((N+1)a {\tl q}) + \gamma^2 \Gamma  \sin((N-2) a {\tl q}) + i \gamma  m r^2 \sqrt{\frac{k (\Omega - 2 \Gamma  \cos(a {\tl q}))}{m r^2}} (\Omega  \sin((N + 1)a {\tl q})  \bigg| \bigg)^2}.\label{GF_qc}
	\eea
\normalsize
The above in Eq.~\ref{GF_qc} can be derived following Eqs.~\ref{e:G1N-square}, \ref{e:detZ}. For relatively large $N$ (and small ${\tl q}$), we may simplify Eq.~\ref{GF_qc} further as
	\bea
	|G_{1,N}|^2 \approx \frac{m^2 r^4 \Gamma^2 \sin^2(a{\tl q})}{\bigg( \left|(2 \Gamma -  \Omega) \left(k mr^2 \Gamma  + \gamma^2  - i \gamma m r^2 \sqrt{\frac{k (\Omega - 2 \Gamma  \cos(a {\tl q}))}{m r^2}} \right) \right|  \sin(N a {\tl q}) \bigg)^2},
	\eea
\end{widetext}
which is $|G_{1,N}|^2$ in Eq.~\ref{e:N-dependence-of-G1N} when $v_1 \neq v_2$. 

We further find that the quantum heat current in the KL chain is independent of $N$ when $v_1 = v_2$. 
We can write the following when $v_1 = v_2$:
\begin{widetext} 
\footnotesize
	\bea 
	|G_{1,N}({\tl q})|^2 &=& \frac{m^2 r^4 \cos^2(a \tl{q}/2)}{\bigg( \bigg| (\gamma^2 - k m r^2 \Gamma) \cos((2N -1) a {\tl q}/2) -\gamma^2 \cos((2N-3) a{\tl q}/2 }\cr
	&& {+ k m r^2 \Gamma \cos((2N + 1) a {\tl q}/2)) + i 2\sqrt{2} \gamma m r^2 \sqrt{\frac{k(\Gamma (1- \cos(a {\tl q})))}{m r^2}} \cos((2N + 1) a {\tl q}/2)) \bigg| \bigg)^2}.
	\eea
\normalsize
\end{widetext}	
For large $N$, this further simplifies into 
	\bea
	|G_{1,N}({\tl q})|^2 \approx \frac{m r^2 \cos^2(a {\tl q}/2)}{16 \gamma^2 k \Gamma \sin^2(a {\tl q}/2) \cos(N a {\tl q})}.
	\eea	
For $|v_1| = |v_2|$, the acoustic branch is gapless, and the low-energy conducting modes are near ${\tl q} \to 0$, for which  $|G_{1,N}({\tl q})|^2$ is almost $N$-independent. Thus, the quantum heat current in Eq.~\ref{HCurrKL1} is system size independent, indicating a ballistic heat transport in the absence of spectral gap at low temperatures. 

\subsection{Classical heat current}
\label{a:Classical-Heat-current-in-Linarized-KL-model}

To evaluate $|G_{1,N}|^2$ in the expression for classical heat current in Eq.~\ref{e:classical-linear-KL-current}, we notice that
	\bea
	|G_{1,N}|^2 = \frac{1}{\tl{z}^2 |\det \tl{Z}|^2}.
	\label{e:G1N-square}
	\eea 
Here, $\tl{z} = k v_1 v_2$ and $\tl{Z} = Z/ \tl{z}$, where $Z$ is the tridiagonal matrix define in Sec.~\ref{s:Heat-conduction-in-an-open-KL-chain}. Thus, the matrix $\tl{Z}$ can be expressed as  
	\bea
	\tl{Z} = \begin{bmatrix}
	z_{L}/{\tl{z}}  & -1 & 0 & \cdots & & 0 \\
	-1  & z/{\tl{z}}  & -1 & \ddots & &\vdots \\
	\vdots & \ddots & \ddots & \ddots &  \ddots &  \vdots \\
	\vdots &  & \ddots & \ddots & \ddots & \vdots\\
	\vdots  & & & -1 & z/{\tl{z}} 	& -1\\
	0 &  \cdots & \cdots & & 	-1 & z_{R}/{\tl{z}}  \\
	\end{bmatrix}.
	\label{e:tl-Z-matrix}
	\eea
Here, the matrix elements have the following form	
	\bea
	\frac{z_L}{\tl{z}} &=& -\frac{m \omega^2 r^2}{k v_1 v_2} + \frac{v_1}{v_2} - \frac{i \gamma_L \omega}{k v_1 v_2} = \frac{z}{\tl z} - \epsilon_L, \cr
	\frac{z_R}{\tl{z}} &=& -\frac{m \omega^2 r^2}{k v_1 v_2} + \frac{v_2}{v_1} - \frac{i \gamma_R \omega}{k v_1 v_2} = \frac{z}{\tl z} - \epsilon_R, \cr
	\frac{z}{\tl{z}} &=& -\frac{m \omega^2 r^2}{k v_1 v_2} + \frac{v_1}{v_2} + \frac{v_2}{v_1},
	\eea 
with $\epsilon_L = (v_2 / v_1) + ( i \gamma_L \omega/k v_1 v_2)$ and $\epsilon_R = (v_1/v_2) + (i \gamma_R \omega/ k v_1 v_2)$. 
\begin{widetext}
In order to find $\det \tl{Z}$, we use the substitution $(z/\tl{z}) = 2 \cos (a {\tl q} )$. If we allow ${\tl q} = 2 \pi n/ Na$, with $n =1, \cdots, N$, then we notice that   
	\bea
	\omega^2(\tl{q}) = \frac{k}{m r^2} (v_1^2 + v_2^2 - 2 v_1 v_2 \cos(a {\tl q})),
	\label{e:omega-in-terms-of-tlq}
	\eea
gives all frequencies in the acoustic and optical branches given in Eq.~\ref{e:dispersion-relation-linear-KL}.  Note that each branches consist of $N/2$ frequencies. The small $\tl{q}$ behaviour of Eq.~\ref{e:omega-in-terms-of-tlq} is the same as that of the acoustic branch in Eq.~\ref{e:dispersion-relation-linear-KL} with the same band gap, when $v_1 \neq v_2$. With this substitution, the determinant can be expressed as 

	\bea
	\det[\tl{Z}_N] = \Delta_N = \frac{A({\tl q}) \sin(N a {\tl q}) + B({\tl q}) \cos(N a {\tl q})}{\sin (a {\tl q})}. 
	\label{e:detZ}		
	\eea 	
Here, 
 	\bea
 	A({\tl q}) = -(\epsilon_L + \epsilon_R) + (1 + \epsilon_L \epsilon_R) \cos(a {\tl q}) \quad \text{and} \quad
 	B({\tl q}) = (1 - \epsilon_L \epsilon_R) \sin (a {\tl q}).
 	\eea	
For $\gamma_L = \gamma_R = \gamma$, we find 	
	\bea
	A({\tl q}) = \frac{-k(v_1^2 + v_2^2)}{k v_1 v_2} + \left( 2 - \frac{ \gamma^2 \omega^2}{(k v_1 v_2)^2} \right) \cos(a {\tl q})  + i \left(\frac{k \gamma \omega (v_1^2 + v_2^2)\cos(a {\tl q})}{(k v_1 v_2)^2}  - \frac{2 \gamma \omega}{k v_1 v_2} \right),
	\eea  	
and
	\bea
	B({\tl q}) = \left( \frac{\gamma^2 \omega^2}{(kv_1 v_2)^2} \right) \sin(a {\tl q})  + i \left( \frac{k \gamma \omega (v_1^2 + v_2^2) \sin(a {\tl q})}{(kv_1 v_2)^2} \right). \quad
	\eea
Thus, the square of the determinant can be expressed as 
	\bea
	|\Delta_N|^2 = \frac{(|A|^2 +|B|^2)(1+r_1 \sin(2N a {\tl q} + \phi))}{2 \sin^2(a {\tl q})},
	\eea	
with $r_1 \cos \phi  = (A B^* + A^* B)/(|A|^2+ |B|^2)$ and $r_1 \sin \phi = (|B|^2 - |A|^2)/ (|A|^2 + |B|^2)$. Now using Eq.~\ref{e:omega-in-terms-of-tlq}, we get the classical heat current in the KL chain:  
	\bea
	\bra J_{\rm KL} \ket^{\rm cl} =  \frac{k_B \Delta T}{\pi} \left( \frac{k}{mr^2} \right)^{3/2} \frac{4 \gamma^2}{k^2 v_1 v_2} \int_{0}^{\pi/a} d{\tl q} \frac{a \sin^3(a {\tl q}) \sqrt{v_1^2 + v_2^2 - 2 v_1 v_2 \cos(a {\tl q})}}{(|A|^2 +|B|^2)(1 + r_1 \sin(2 N a {\tl q} + \phi))}.
	\eea	
In the large $N$ limit, using the identity
	\bea
	\lim_{N \to \infty} \int_{0}^{\pi}  d{\tl q} \frac{g_1({\tl q})}{1+ g_2({\tl q}) \sin(N {\tl q})} = \int_{0}^{\pi} d{\tl q} \frac{g_1({\tl q})}{(1- g_2({\tl q})^2)^{1/2}},
	\eea	
along with the identifications $g_2({\tl q}) = r_1$, and 
	\bea
	g_1({\tl q}) = \frac{a \sin^3(a {\tl q}) \sqrt{v_1^2 + v_2^2 - 2 v_1 v_2 \cos (a {\tl q})}}{(|A|^2 +|B|^2)},
	\eea
and noticing that
	\bea
	(1- g_2({\tl q})^2)^{1/2} = \frac{2(|a_1 b_2 - a_2 b_1|)}{|A|^2 + |B|^2},	
	\eea
we get the classical heat current 
	\bea
	\bra J_{\rm KL} \ket^{\rm cl} =	\frac{k_B \Delta T}{\pi} \left( \frac{k}{mr^2} \right)^{3/2} \frac{2 \gamma^2}{k^2 v_1 v_2} \int_{0}^{\pi/a} d{\tl q} \frac{a \sin^3(a {\tl q}) \sqrt{v_1^2 + v_2^2 - 2 v_1 v_2 \cos(a {\tl q}) }}{(|a_1 b_2 - a_2 b_1|)}.
	\eea
Here, $A = a_1 + i a_2$ and $B = b_1 + i b_2$. Thus, we have
	\bea
	|a_1 b_2 - a_2 b_1| = \frac{\gamma  \left(\frac{k}{m r^2}(v_1^2 + v_2^2 -2 v_1 v_2 \cos (a {\tl q})) \right)^{3/2} (v_1 v_2 (k m r^2(v_1^2 + v_2^2) - 2 \gamma^2) + 2 (v_1^2 + v_2^2) \gamma^2 \cos(a {\tl q})) \sin (a {\tl q})}{k^3  v_1^4 v_2^4}. \qquad \quad
	\eea		 
Applying this in the above expression, we get the classical heat current in terms of the system parameters as 
	\bea
	\bra J_{\rm KL} \ket^{\rm cl} = \frac{k_B \Delta T}{\pi} \int_{0}^{\pi/a} \frac{d{\tl q} \, 2 a \sin^2(a {\tl q}) \gamma k \Gamma^3}{(\Omega - 2 \Gamma \cos(a {\tl q}))(\Gamma(k m r^2 \Omega - 2 \gamma^2) + 2 \Omega \gamma^2 \cos(a {\tl q}))}.
	\eea 
Here, $\Omega = v_1^2 + v_2^2$ and $\Gamma = v_1 v_2$. By a change of variable $x = \cos(a {\tl q})$, we re-express the heat current as 
	\bea
	\bra J_{\rm KL} \ket^{\rm cl} = \frac{k_B \Delta T}{\pi} \int_{-1}^{1} \frac{2 \gamma k \Gamma^3 \sqrt{1-x^2} \ dx}{(\Omega - 2 \Gamma x)(\Gamma( k m r^2 \Omega - 2 \gamma^2) + 2 \Omega \gamma^2 x)}.
	\eea
We thus get the analytical form of the classical heat current in Eq.~\ref{e:classical-current-KL}, when $2\Omega \gamma^2 < \Gamma (k m r^2 \Omega - 2 \gamma^2)$. In the special case of $\bar{\theta} = 0$ (i.e., $v_1 = v_2$ implying $\Omega = 2 \Gamma$), this formula can be further simplified to 
	\bea
	\bra J_{\rm KL} \ket^{\rm cl}  =   \frac{ k_B \Delta T\Gamma  k \left(4 \gamma ^2+2 \left(\gamma ^2-\Gamma  k m r^2\right) \left(\sqrt{1-\frac{4 \gamma ^4}{\left(\gamma ^2-\Gamma  k m r^2\right)^2}}-1\right)\right)}{8 \left(\gamma^3 + \gamma  \Gamma  k m r^2\right)}.
	\eea

\end{widetext}
\bibliography{bibliographyKL}

\end{document}